\documentclass[conference]{IEEEtran}
\IEEEoverridecommandlockouts
\usepackage{cite}
\usepackage{amsmath,amssymb,amsfonts}
\usepackage{algorithmic}
\usepackage{algorithm}
\usepackage{graphicx}
\usepackage{textcomp}
\usepackage{diagbox}
\usepackage{xcolor}
\usepackage{bm}
\usepackage{bbm} 
\usepackage{placeins}
\usepackage{hyperref}

\usepackage[top=0.77in, bottom=1.05in, left=1.02in, right=1.02in]{geometry}

\def\BibTeX{{\rm B\kern-.05em{\sc i\kern-.025em b}\kern-.08em
    T\kern-.1667em\lower.7ex\hbox{E}\kern-.125emX}}
\begin{document}

\newcommand{\ceil}[1]{\left\lceil #1 \right\rceil}

\newcommand{\one}{\mathbbm{1}} 
\title{Self-Supervised Transformer-based Contrastive Learning for Intrusion Detection Systems
\thanks{This work has received co-funding from the Smart Networks and Services Joint Undertaking (SNS JU) under the European Union’s Horizon Europe research and innovation programme, in the frame of the NATWORK project (Net-Zero self-adaptive activation of distributed self-resilient augmented services) under Grant Agreement No 101139285.}
}

\author{\IEEEauthorblockN{ Ippokratis Koukoulis, Ilias Syrigos, and Thanasis Korakis}  
\IEEEauthorblockA{\textit{Dept. of Electrical and Computer Engineering, University of Thessaly, Greece}\\
\textit{Centre for Research and Technology Hellas, CERTH, Greece}\\
Email: ikoukoulis@uth.gr, ilsirigo@uth.gr, korakis@uth.gr
\\[-2.5ex]
 }}

\maketitle

\begin{abstract}

As the digital landscape becomes more interconnected, the frequency and severity of zero-day attacks, have significantly increased, leading to an urgent need for innovative Intrusion Detection Systems (IDS). Machine Learning-based IDS that learn from the network traffic characteristics and can discern attack patterns from benign traffic offer an advanced solution to traditional signature-based IDS. However, they heavily rely on labeled datasets, and their ability to generalize when encountering unseen traffic patterns remains a challenge.
This paper proposes a novel self-supervised contrastive learning approach based on transformer encoders, specifically tailored for generalizable intrusion detection on raw packet sequences. Our proposed learning scheme employs a packet-level data augmentation strategy combined with a transformer-based architecture to extract and generate meaningful representations of traffic flows. Unlike traditional methods reliant on handcrafted statistical features (NetFlow), our approach automatically learns comprehensive packet sequence representations, significantly enhancing performance in anomaly identification tasks and supervised learning for intrusion detection.
Our transformer-based framework exhibits better performance in comparison to existing NetFlow self-supervised methods. Specifically, we achieve up to a 3\% higher AUC in anomaly detection for intra-dataset evaluation and up to 20\% higher AUC scores in inter-dataset evaluation. Moreover, our model provides a strong baseline for supervised intrusion detection with limited labeled data, exhibiting an improvement over self-supervised NetFlow models of up to 1.5\% AUC when pretrained and evaluated on the same dataset. Additionally, we show the adaptability of our pretrained model when fine-tuned across different datasets, demonstrating strong performance even when lacking benign data from the target domain.

\end{abstract}


\section{Introduction}


Digital connectivity is significantly expanding day by day and we now live in a highly interconnected world. At the same time, well-known cyber threats are still being encountered worldwide, there are new emerging ones that are often unknown to existing defense mechanisms and cybersecurity experts, consequently going undetected to form zero-day attacks. According to Rapid7's Attack Intelligence Report \cite{rapid7Report2024}, 2023 was the year that showed a higher percentage of zero-day vulnerabilities than any previous year, with every such breach resulting in significant operational and financial costs, as highlighted by IBM 2024 Data Breach Report \cite{ibmBreach2024}, which indicates that the average cost of a data breach stands at 4.88 million dollars. Therefore, there is an urgent need for intelligent and innovative approaches that extend the traditional threat detection mechanisms to offer accurate and effective zero-day threat detection.

Traditional intrusion detection systems are unable to identify new threats if there does not already exist a known signature for them, while anomaly detection systems often fail to distinguish between malicious and legitimate anomalies, which leads to a lot of false positives \cite{zohrevand2019should}. To overcome these challenges, many approaches have been developed that employ Machine Learning and Deep Learning methods to recognize abnormal traffic patterns, thereby enabling them identify known and zero-day attacks more effectively. Although these supervised methods are accurate, they are limited because they require a large and carefully labeled training dataset. This limits their capacity to generalize to various types of traffic and attacks and turns them inefficient, as it renders them overly reliant on the manual labor required to label the traffic flows.

Self-supervised learning (SSL) techniques, such as transformer-based architectures and large language models (LLMs), offer great potential for overcoming the limitations and shortcomings of supervised learning, as highlighted by recent advancements in Deep Learning, especially in the domains of computer vision and natural language processing (NLP). SSL techniques are able to learn useful representations from unlabeled data that can be leveraged for a variety of downstream tasks, such as classification. In addition, contrastive learning, a process that has been employed to extract representations from images \cite{chen2020simple} as well as from text \cite{gao2021simcse}, is superior at learning representations by maximizing the similarity between related data points and minimizing dissimilar ones. Consequently, contrastive learning models are significantly more generalizable and robust, when encountering new and previously unseen traffic patterns. This paradigm has also been recently adapted in the domain of tabular data with conventional neural network architectures \cite{bahri2021scarf} and transformer architectures \cite{somepalli2021saint}.

Related works that are based on self-supervised contrastive learning \cite{golchin2024sscl} derive insights from flow-level statistical features such as traffic volume and packet frequency. These features are capable of significant predictive capabilities. Nevertheless, they are unable to encapsulate packet-based specific information and frequently require manual selection based on domain expertise and traffic characteristics, resulting in a reliance on handcrafted features that significantly limits their adaptability to a variety of network environments. Additionally, the ML models that are trained on a fixed statistical distribution may be unable to detect novel attack strategies as network architectures and cyber threats continue to evolve. In order to address these limitations, it is essential to implement a more advanced methodology that is capable of automatically extracting meaningful representations from network traffic sequences.

In this paper, we propose a novel transformer-based framework that can extract a flow representation from a sequence of packets. To achieve this, we train our model using a contrastive learning process, where the model learns to
minimize the distance between an original flow packet sequence and an augmented one, while maximizing the distance with other samples. To create augmented packet sequences, we employ a process that allows us to mix packets of the original flow with that of another one, to create a new sequence that is similar to the original. The source code for the model implementation is provided here: \mbox{\url{https://github.com/koukipp/contrastive_transformers_ids}}.

The main contributions of our paper are the following:
\begin{itemize}
    \item The development of a novel transformer-based framework pretrained on unlabeled data traffic that can extract useful flow representations directly from raw packet sequences, reducing the need for manual extraction of statistical features from flows.
    \item The proposal of a simple augmentation process that enables effective contrastive learning on sequences of packets.
    \item The evaluation of the extracted representations in an unsupervised setting to detect anomalous flows and demonstrate our framework's ability to recognize novel attacks across multiple datasets.
    \item The demonstration of the effectiveness of our pretraining procedure to enhance the generalization of intrusion detection in supervised settings, where the model is finetuned with a small amount of samples.
\end{itemize}

\section{Related Work}

\subsection{Transformer-based models for network traffic}

In the context of sequential modeling for network traffic \cite{dietmuller2022new}, transformer \cite{vaswani2017attention} architectures have gained traction. This is due to the fact that general models can be fine-tuned on a wide range of downstream tasks using labeled data, including intrusion detection, after being pretrained on unlabeled network traffic.  In \cite{guthula2023netfound}, the authors propose a foundation model that facilitates pretraining on raw packet data and fine-tuning on specific security-related tasks. In this work, the packet trace of each flow is tokenized by splitting the header and a small portion of the payload into tokens and the resulting sequence is then fed into the model to classify each flow. Similarly, the authors in \cite{han2023network} employ n-gram frequency to tokenize the entire payload of the packet. Transformer-based architectures have also been implemented in flow-based intrusion detection datasets to classify each flow using its statistical features as tokens \cite{wu2022rtids} or by using temporally related flows as tokens for a sequence \cite{manocchio2024flowtransformer,nguyen2023method}. Additionally, they have been used in packet-based datasets to classify each packet individually by using the headers of each packet as tokens \cite{ferrag2024revolutionizing}.
 Furthermore, transformer-based models have been employed in unsupervised learning schemes \cite{wang2023robust} to train the model exclusively with benign flows in the context of anomaly detection, as well as in semi-supervised schemes \cite{li2022extreme} where the model is trained with a very small proportion of labeled data.


Transformer-based architectures have also been employed in the broader task of traffic classification.  Modified language models, such as BERT \cite{lin2022bert} or GPT \cite{meng2023netgpt}, have been employed to identify encrypted traffic flows that originate from a variety of applications. These models use the headers of a sequence of packets as tokens to construct a sentence. Additionally, other proposals have taken into account inter-flow temporal relationships to identify applications \cite{zhao2022mt}.


In addition, transformer-based models have been employed in several works to conduct real-time traffic classification or intrusion detection. The authors in \cite{tan2019neural} have employed a modified transformer model to identify attacks in real-time. Each token in the sequence is a statistical feature vector that describes the aggregate traffic in the network for a specific time slot. In contrast, the authors in \cite{babaria2021flowformers} collect statistical feature vectors for each time slot per flow as tokens to feed into a transformer-based model to perform per-flow traffic classification. Transformers have also been used in conjunction with reinforcement learning \cite{chen2023real} on packet sequences of flows to determine a lower bound on which packet of the sequence the agent can confidently make a prediction, thereby balancing the tradeoff between accuracy and timeliness of the prediction.

\subsection{Self-Supervised learning for intrusion detection}
In \cite{wang2023robust} the authors propose a self-supervised scheme using a transformer architecture which combines contrastive learning with mask reconstruction within a sequence of flow statistical vectors, to provide robust intrusion detection in unsupervised settings.
In \cite{yue2022contrastive} the authors propose a a contrastive learning approach combined with supervised learning on sequences of packets using a variety of CNN and LSTM models. The augmentation method employed here is to mask a small number of packets in a sequence to create an augmented packet sequence. However, this augmentation method does not create a challenging contrastive learning process since the augmented sample barely differs from the original.
Similarly in \cite{wang2021network} the authors propose an augmentation process that treats each packet sequence as an image and apply relevant augmentations such as horizontal/vertical flip, random cropping, and shuffling. Additionally, similar masking augmenting techniques have been shown to be inefficient compared to techniques that alter the contents of a sample \cite{yun2019cutmix}. In \cite{caville2022anomal} the authors investigate the use of Graph Neural Networks (GNNs) for self-supervised intrusion and anomaly detection in computer networks. The authors in \cite{golchin2024sscl} propose a framework that leverages contrastive learning on flow-based statistics. The authors employ an augmentation process adopted for tabular data \cite{bahri2021scarf} that generates new samples by randomly replacing features in the original sample from the empirical marginal distribution of each feature.
\section{Dataset preprocessing}



To evaluate the real-time performance of our model, we developed a straightforward processing pipeline that captures information regarding the initial patterns that emerge in a flow. A flow is defined as a 5-tuple comprising the IP source address, IP destination address, source port, destination port, and protocol, while packets lacking an IP header are eliminated from the packet traces. For each packet trace, we generate two distinct datasets; a packet-sequence dataset comprising a sequence of packets for each flow, alongside a flow-statistics dataset, henceforth referred to as NetFlow, which contains aggregate statistics for each flow analogous to flow-based datasets found in the literature \cite{sharafaldin2018toward} \cite{moustafa2015unsw}.

 We utilized a modified version of CICFlowmeter\cite{engelen2021troubleshooting}, a network traffic analysis tool, to generate the NetFlow dataset, which produces a statistical feature vector for each flow comprising 43 unique features. We capture packets until the flow concludes or until specific time or packet count thresholds are attained. Specifically, we configure the flow timeout to 120 seconds and capture only the initial 32 packets of each flow. This guarantees that each flow is distinct and not an element of a larger flow. Consequently, packets arriving beyond this time threshold are excluded from the calculation of the flow statistics. Similarly, We developed a script to generate the packet-sequence dataset, which implements the same pre-processing logic and produces a truncated sequence of packets received prior to the expiration of the flow timeout limit.

\begin{table}[ht]
\centering
\caption{Dataset features}
\label{table:2}
\begin{tabular}{|c|c|} 
 \hline
 \textbf{Packet Sequence} & \textbf{Netflows} \\
 \hline\hline
  IP Protocol & IP Protocol \\ 
  
  Packet Length & Mean, Max, Min, Std, Total Packet size  \\

  TCP flags & TCP flag counts  \\

  Inter Arrival Time & Mean, Max, Min, Std, Inter Arrival Time \\
  
  Direction & Flow Duration \\
  
    & Packet counts \\

 \hline
\end{tabular}
\end{table}
\hspace{5mm}

We eliminate all data from the NetFlow and packet-sequence datasets that could potentially identify an attacking host, including IP addresses and port numbers.  We utilize, however, this information to develop a new feature termed 'direction', which indicates the direction of each packet in the sequence between the source and destination. For the packet-sequence dataset, we specifically selected packet headers outlined in Table \ref{table:2} that correspond to the relevant features of the NetFlow dataset. We meticulously ensured that none of the supplementary features enable shortcuts in the learning process, as fields in packet headers, like TTL, have been shown to convey information regarding the distance from the source \cite{holland2021new}.

 In our tokenization process, we establish a vocabulary size of 65538, comprising two special tokens: the [PAD] token, utilized to standardize the sequence length of each sample in a batch to the maximum sequence length; the [CLS] token, designed to generate a representation of the flow. We encode each packet header as a 4-byte unsigned integer, with each token ranging from 0 to 65535.

\begin{figure*}[ht]
\centering
\includegraphics[width=0.7\textwidth]{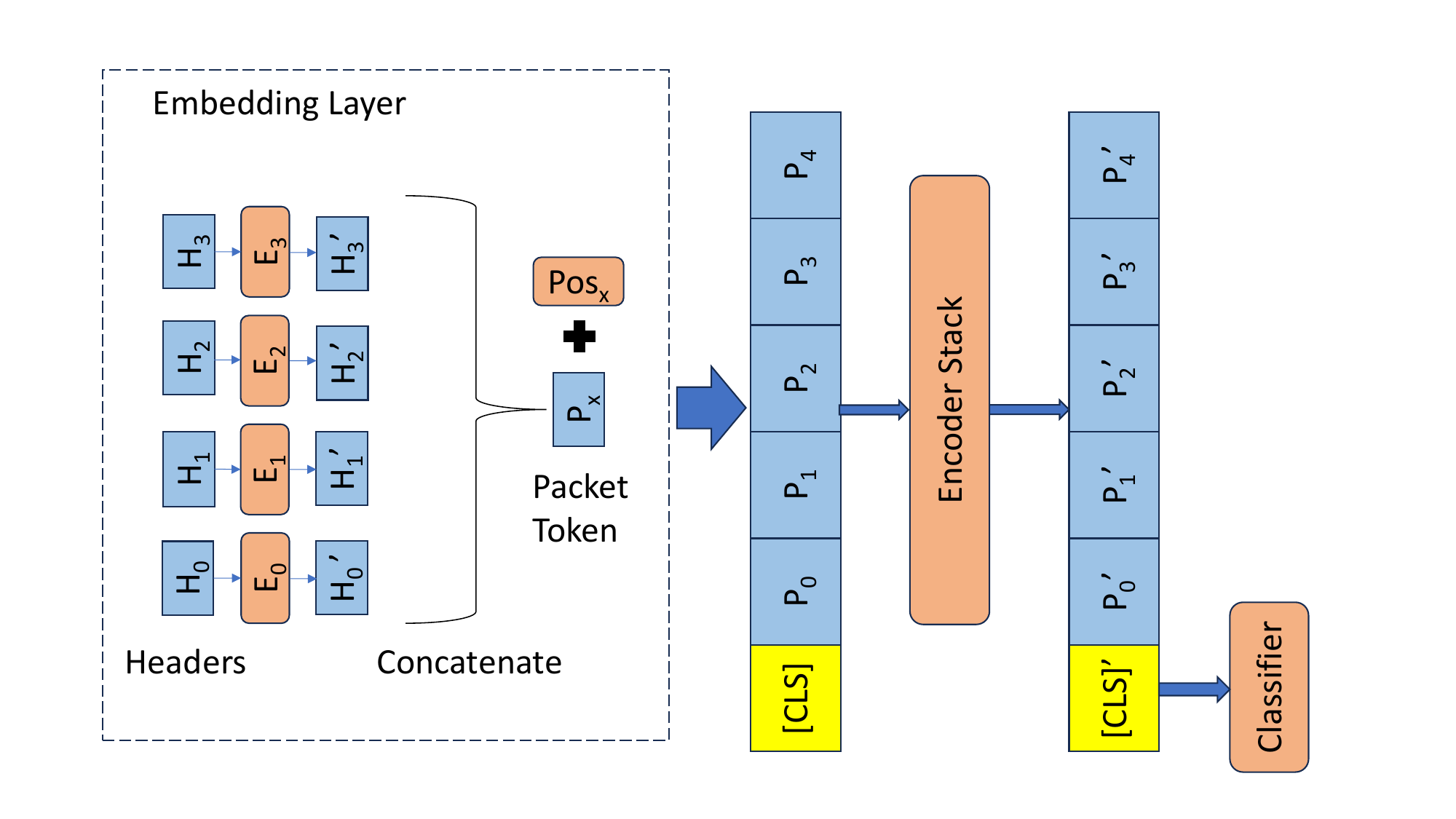}
\vspace{-4mm}
\caption{Overall architecture of the transformer-based model}
\label{fig:arch}
\end{figure*} 

\section{Overview of model architecture}
We employed an architecture based on the BERT \cite{devlin2018bert} language model to facilitate intrusion detection classification with our model.
We selected the BERT transformer encoder stack as the basis of our model architecture due to its inherent suitability for classification tasks, unlike architectures such as GPT\cite{radford2018improving}, which are primarily designed for generative tasks. BERT utilizes bidirectional self-attention to extract comprehensive contextual information from a sentence, facilitating tasks associated with language comprehension, including question answering and language inference.

\subsection{Packet-based token embedding}
Processing packets as a sequence of tokens requires an efficient embedding procedure that maintains the semantic integrity of the data. Related studies have employed various approaches in which unique tokens are generated for each byte or header of a packet. Nonetheless, these approaches possess several drawbacks. Firstly, the large amount of generated tokens drastically impacts inference and training duration, as attention computations in transformer models scale quadratically with sequence length. This renders inference on extended sequences of packets impractical for real-time applications. Moreover, the presence of distinct headers or bytes as tokens complicates the learning process for any task, as the model must independently recognize which tokens are part of the same packet. Finally, in contrast to conventional language-derived tokens, packet headers encompass a wide variety of features, both categorical and numerical, which cannot be effectively represented by tokens using a single embedding layer approach.
To address these issues, we implemented an efficient packet-based token embedding scheme that generates a single token for each packet in a sequence. This approach significantly reduces the sequence length, enabling the processing of more tokens within reasonable timeframes. Figure \ref{fig:arch} illustrates the pipeline of our tokenization and embedding process.

To generate a packet-token, we take the value of each header $H_{x} \in \mathbb{R}$ to initially produce individual packet-header tokens for the selected packet headers. The tokens are forwarded to an embedding layer to obtain an embedding vector that represents each packet header. For each header, we employ a unique embedding function $E(x)$ to generate the header token embedding $H_{x}' \in \mathbb{R}^{d_{h}}$, where $d_{h}$ denotes the dimension of the header token embedding. We employ distinct embedding functions, unlike the conventional single embedding layer utilized in transformer models for NLP tasks, since each header token derives from a different domain. Specifically to generate tokens for categorical features such as the direction and the flags of each packet, we use a typical embedding layer that maps each discrete value to a floating point vector with trainable weights. To generate a header token embedding for numerical values and preserve the ordinality of these features, such as the inter-arrival time between packets and the packet size, we employ a single layer projections as the embedding functions $E(x)$ for each numerical header feature. Ultimately, the header token embeddings are concatenated into a single packet-token \( P_{x} \in \mathbb{R}^{d} \) via a linear layer \( L \in \mathbb{R}^{n_{h}*d_{h} \times d} \), where \( d \) represents the embedding dimension of the packet token, thereby forming the input sequence for the BERT encoder stack. We utilize a positional embedding layer $Pos \in \mathbb{R}^{L_{max} \times d}$ to convey information regarding the position of each packet token within the sequence, where $L_{max}$ denotes the maximum sequence length, and this layer is trained concurrently with the remainder of the model. The representation of each position $Pos_{x}$ is directly incorporated into each token $P_{x}$. Alongside the packet tokens, we prepend a special [CLS] token at the start of the sequence, which serves as the model's output.

\subsection{Encoder architecture}
The encoder layer of our model is composed of a series of stacked Transformer encoders, which apply self-attention on a sequence of tokens in order to capture correlations between the tokens. Each transformer encoder is comprised of a multi-head self-attention layer and a fully connected feed-forward network. The input to the encoder stack is the packet-based token sequence. Each token of the packet sequence is linearly projected to multiple attention heads that are separated into a query $Q$, key $K$ and $V$ value vectors with dimension $d_{k}$, which are used to calculate the scaled dot product attention as defined in Equation \ref{attention} for each head.  The output of each attention head is then concatenated and linearly projected to a vector that has the same shape as the initial input sequence.
Multi-headed attention allows the transformer to attend to different characteristics of the packet feature space allowing the model to find correlations between the different packet headers in each packet of the sequence.

\begin{equation} \label{attention}
\text{Attention}(Q, K, V) = \text{softmax}\left(\frac{QK^\mathrm{T}}{\sqrt{d_k}}\right)V
\end{equation}

For our architecture we selected a model consisting of 4 layers of stacked transformer encoders with an embedding dimension $d$ = 256, with 4 attention heads. While this is a small-scale model compared to the regular size of other transformer encoder models \cite{devlin2018bert} used for NLP tasks, it is sufficient to capture the characteristics of a packet sequence, while keeping the processing time for inference low.

\subsection{Projection head}
To train our model using the contrastive learning objective we use a projection layer on the output of the [CLS] token from the transformer encoder stack. The projection layer comprises a multi-layer perceptron (MLP) featuring a single hidden layer that employs the ReLU activation function. During inference, the projection head is discarded and the output of the [CLS] token is utilized to evaluate the quality of the flow representations.

\subsection{Contrastive learning}

The objective of the contrastive learning task is to acquire meaningful representations by minimizing the distance between similar samples in the embedding space while maximizing the distance between dissimilar samples. To accomplish this, we employ the NT-Xent (Normalized Temperature-scaled Cross-Entropy) loss function \cite{chen2020simple}, as delineated in the Equation \ref{contrastive_loss}, where $(\mathbf{z}_i, \mathbf{z}_j)$ represent instances of similar "positive" pairs, while $(\mathbf{z}_i, \mathbf{z}_k)$ denote instances of dissimilar "negative" pairs, which, in our self-supervised context, encompass all other samples within a batch. The temperature parameter $\tau$ regulates the sensitivity of the loss function by adjusting the cosine distance. Lower values of $\tau$ increase the penalty for discrepancies within a pair.

\begin{equation} \label{contrastive_loss}
\mathcal{L}_{i,j} = -\log \frac{\exp\left(\operatorname{sim}(\mathbf{z}_i, \mathbf{z}_j) / \tau\right)}
{\sum\limits_{k=1}^{2N} \mathbf{1}_{[k \neq i]} \exp\left(\operatorname{sim}(\mathbf{z}_i, \mathbf{z}_k) / \tau\right)}
\end{equation}

\begin{figure}[t]
\centering
\includegraphics[width=\columnwidth]{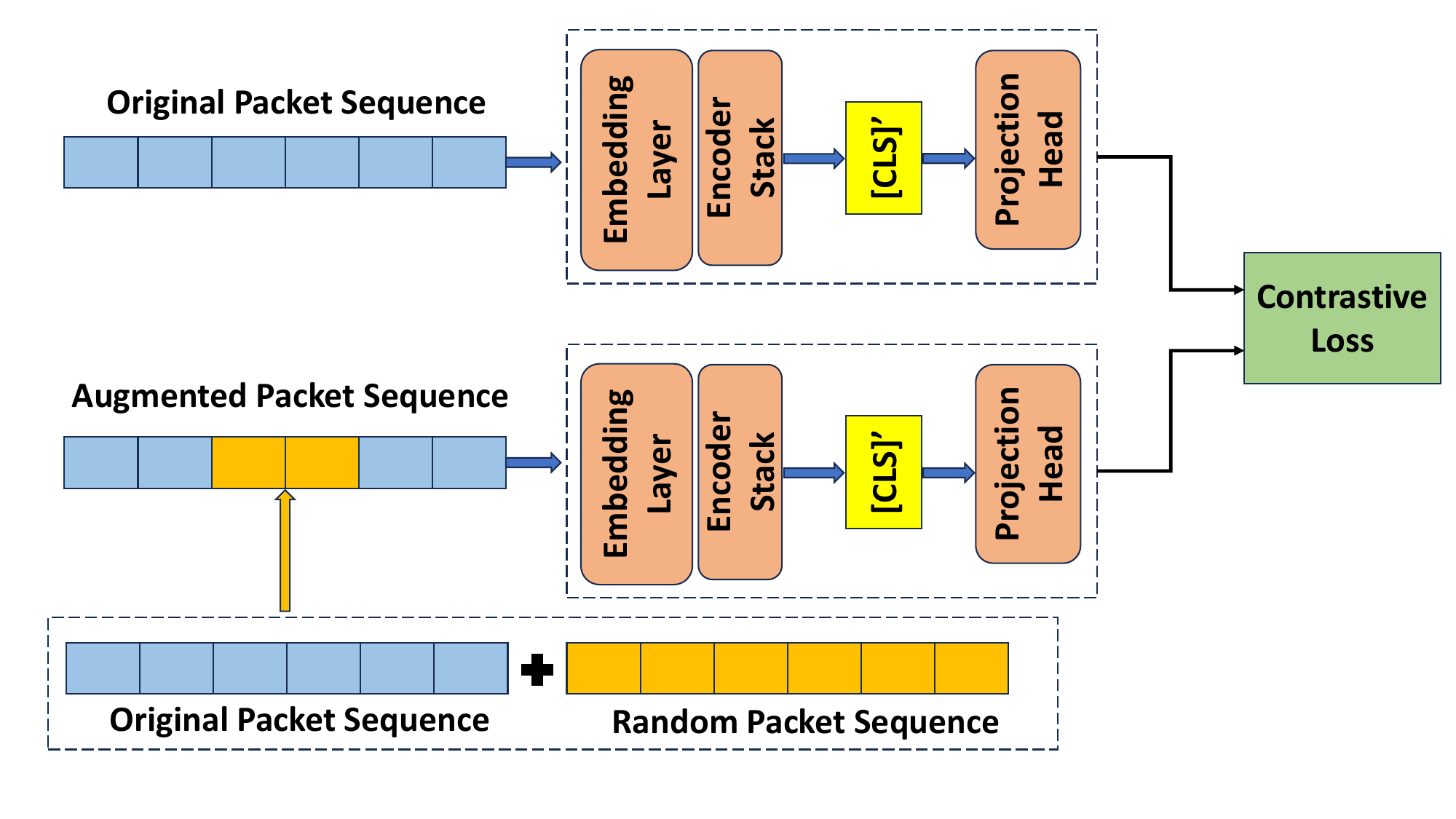}
\caption{Overview of the augmentation and contrastive learning procedure}
\label{fig:arch}
\end{figure} 

\subsection{Augmentation process}
Contrastive learning can be either supervised, using labels to identify similar and dissimilar samples, or self-supervised, wherein an augmentation process generates modified views of existing samples to provide examples of similar instances for the training process. We devised a straightforward procedure for generating augmented views of our samples, as described in Algorithm 1, of which we provide an overview in Figure \ref{fig:arch}, in conjunction with the training process. To generate positive pairs of flow packet sequences for each sample, we select another sample of equivalent size from the training dataset for the augmentation process. Subsequently, from the chosen sample, we extract a random segment of contiguous packets and replace them in the batch sample. Similar processes have been proven effective in guiding a model to attend to less discriminative parts of an input \cite{yun2019cutmix} leading to better generalization. 

\begin{algorithm}
\caption{Contrastive training process}\label{alg:cap}
\begin{algorithmic}
    \STATE \textbf{input:} unlabeled training data $\mathcal{X} \subseteq \mathbb{R}^{M}$, batch size $N$, temperature $\tau$, encoder $f$, projection head $g$, augmentation ratio $\lambda$.
    \FOR{sampled mini-batch $\left\{x^{(i)}\right\}_{i=1}^N \subseteq \mathcal{X}$}
    \STATE \textbf{for all} $i\in \{1, \ldots, N\}$ \textbf{do}
        \STATE $~~~~$draw sample $u \!\sim\! \mathcal{X}$
        \STATE $~~~~$\textcolor{gray}{\# Choose the beginning of the patch} 
        \STATE $~~~~$${\bm a} \!\sim\!$ Uniform $(0,L * (1 - \lambda))$
        \STATE $~~~~$\textcolor{gray}{\# Create augmented sample} 
        \STATE $~~~~$$\tilde{\bm x}^{(i)}_{j} = \bm u_{j}$ if $j \in (\alpha,\alpha + \lambda * L)$
        \STATE $~~~~$$\tilde{\bm x}^{(i)}_{j} = \bm x^{(i)}_{j}$ otherwise
        \STATE $~~~~$\textcolor{gray}{\# Get embeddings} 
        \STATE $~~~~$$\bm z^{(i)} = g(f({\bm x}^{(i)}))$ 
        \STATE $~~~~$$\tilde{\bm z}^{(i)} = g(f(\tilde{\bm x}^{(i)}))$  
    \STATE \textbf{end for}
    \STATE \textbf{for all} $i\in\{1, \ldots, N\}$ and $j\in\{1, \dots, N\}$ \textbf{do}
    \STATE $~~~~$ $s_{i,j} = \bm z^{i\top} \tilde{\bm z}^j / (\lVert\bm z^i\rVert \lVert\tilde{\bm z}^j\rVert)$ \textcolor{gray}{~~~~~~~~\# pairwise similarity}\\
    \STATE \textbf{end for}
    \STATE \textbf{define} $\ell(i, j)$ \textbf{as}
    \STATE \parbox[t]{\linewidth}{$\ell(i, j) \!=\! -\log \frac{\exp(s_{i,j}/\tau)}{\sum_{k=1}^{2N} \one{k \neq i}\exp(s_{i, k}/\tau)}$}
    \STATE update networks $f$ and $g$ to minimize $\mathcal{L}$
    \ENDFOR
    \STATE \textbf{return} encoder network $f(\cdot)$, and throw away $g(\cdot)$
\end{algorithmic}
\end{algorithm}

\subsection{Fine-tuning}
Once the model has been pretrained to learn the relationships between packet headers and packets within a flow sequence, it can serve as a basis for training a fine-tuned model for intrusion detection tasks. We employ a two-layer MLP classifier that utilizes the output of the [CLS] token from the final hidden layer of our model to generate a prediction regarding the class of the flow sequence.  During the fine-tuning process, we train both the novel classifier and the pretrained model. This is done to ensure that the weights of the transformer encoder stack are also trained on malicious data to provide an improved representation of the flow in the [CLS] token output for the classifier.

\section{Evaluation}
To evaluate our framework we measure the performance of our pretrained model across unsupervised and supervised tasks to demonstrate its capacity to enable detection of unseen attacks as well as adapt to traffic data from various domains.
We evaluate our model using the packet traces of various datasets as cited below:
\begin{itemize}
\item CICIDS2017 \cite{sharafaldin2018toward}. This dataset contains a diverse set of attacks and benign traffic that were gathered within a five-day period. More specifically, it includes traffic for 14 distinct
attacks including Brute Force Attacks (FTP-Patator, SSH-Patator), DoS attacks (Hulk, Golden-Eye, Slowloris, Slowhttptest, Heartbleed), Web Attacks (Brute
Force, XSS, and SQL Injection), Infiltration Attacks, Botnet, DDoS and PortScan. We also relabeled the attacks from the original dataset according to \cite{engelen2021troubleshooting}\cite{liu2022error} as a lot of the samples were mislabeled. For the Web Attack, Infiltration and Botnet classes, particularly, we only consider flows that actually carry payload during the published attack time frame to be malicious.
\item UNSW-NB15 \cite{moustafa2015unsw}. This dataset contains benign and malicious traffic generated by a traffic simulation hardware and provides a hybrid of real modern normal activities and synthetic contemporary attack behaviors. The wide range of attacks in this dataset includes Fuzzers, Analysis, Backdoors, DoS, Exploits, Generic, Reconnaissance, Shellcode and Worms.
\item CTU-13 \cite{garcia2014empirical}. This dataset, which was captured in a university network, includes a combination of malicious botnet activity, normal traffic originating from known hosts, and background traffic originating from unknown hosts. Specifically, the dataset comprises 13 different captures each containing malicious traffic traces generated by different malware such as Neris, Rbot, Virut, Menti, Sogou, Murlo and NSIS.ay
\item CIC-DDoS2019 \cite{sharafaldin2019developing}. This dataset contains an extended set of Distributed Denial of Service (DDoS) attacks along with a small subset of benign traffic, all of which were collected within a two-day period. The DDoS attacks in this dataset were generated using a variety of methods to increase the impact of the attack, including reflection, and the exploitation of legitimate third-party components to conceal the attacker's identity. Reflection-based attacks include DDoS attacks against a variety of protocols such as MSSQL, SSDP, DNS, LDAP, NETBIOS, SNMP, NETBIOS, CharGen, NTP, TFTP. Exploitation attacks, on the other hand, primarily employ SYN and UDP Flood.
\end{itemize}

\begin{table}[t] 
    \centering
    \caption{Unsupervised evaluation for contrastive transformers on packet sequences}
    \label{table:unsupervised_transformer}
    \resizebox{\columnwidth}{!}{ 
    \begin{tabular}{|c|c|c|c|c|} 
     \hline
      \textbf{\backslashbox{Train}{Test}} & \textbf{CICIDS} & \textbf{UNSW-NB}  & \textbf{CTU}  & \textbf{CICDDOS} \\ [0.3ex] 
     \hline
     \textbf{CICIDS} & \textbf{99\%} & \textbf{79\%} & 63\% & \textbf{62\%}\\
     \hline
     \textbf{UNSW-NB} & \textbf{84\%} & 95\% & \textbf{80\%} & \textbf{55\%} \\ 
     \hline
     \textbf{CTU} & \textbf{94\%} & \textbf{75\%} & 96\% & \textbf{60\%}\\ 
     \hline
     \textbf{CICDDOS} & \textbf{92\%} & \textbf{60\%} & \textbf{55\%} & 93\% \\
     \hline
    \end{tabular}
    }
\end{table}

\begin{table}[t] 
    \centering
    \caption{Unsupervised evaluation for contrastive DNN on Netflows}
    \label{table:unsupervised_dnn}
    \resizebox{\columnwidth}{!}{ 
    \begin{tabular}{|c|c|c|c|c|} 
     \hline
      \textbf{\backslashbox{Train}{Test}} & \textbf{CICIDS} & \textbf{UNSW-NB}  & \textbf{CTU}  & \textbf{CICDDOS} \\ [0.3ex] 
     \hline
     \textbf{CICIDS} & 97\% & 77\% & 53\% & 58\%\\
     \hline
     \textbf{UNSW-NB} & 67\% & 93\% & 60\% & \textbf{55\%}\\ 
     \hline
     \textbf{CTU} & 92\% & 55\% & 93\% & 58\%\\ 
     \hline
     \textbf{CICDDOS} & 91\% & 53\% & 53\% & 91\% \\
     \hline
    \end{tabular}
    }
    \label{tab:bigtable}
\end{table}

\subsection{Evaluation setup}
For the implementation of our model we used the Pytorch framework, along with an RTX 4070 12GB GPU for the training and inference process.
To train our model for the self-supervised task we used a batch size of 128 and the AdamW optimizer with a learning rate of $5e^{-5}$ for 1 epoch on the training subset of each dataset. For our contrastive loss function, the temperature parameter $\tau$ was set to 0.5, and the augmentation ratio $\lambda$ to 0.4. Additionally we applied a 0.1 dropout probability in the transformer encoder stack.
For the training dataset used during self-supervised training, both for intra-dataset and inter-dataset evaluation, we select 60\% of the unlabeled benign traffic from each dataset. The remaining benign flows and all malicious flows included in the dataset are utilized for the testing dataset in the unsupervised evaluation. For supervised training we use the same percentage of benign traffic for the training process. However, we also add an additional 60\% of the total malicious traffic from the same dataset, while the remaining flows are used for the testing dataset in the supervised evaluation.

To compare the performance of our model with similar self-supervised schemes on NetFlows, we employ a 4-layer DNN and an embedding dimension of 256 as the encoder, along with a 2-layer MLP as the projection head (or as the classification head when supervised finetuning or evaluation takes place). This DNN is trained with a similar contrastive learning process. NetFlow features are corrupted randomly by replacing them from the empirical marginal distribution of each feature in the augmentation process for the DNN model \cite{bahri2021scarf}\cite{golchin2024sscl}. The augmentation ratio that we employed for our model remains unchanged. We compare the performance of each model both in the supervised and unsupervised cases using the Area Under the Receiver Operating Characteristic Curve (AUC-ROC) score. The AUC-ROC score is a relative comparison metric that evaluates the ability of each model to distinguish between positive (malicious) and negative (benign) instances. It assesses the true positive rate and false positive ratio across multiple similarity thresholds for the unsupervised evaluation or probability thresholds for the supervised evaluation.

\subsection{Unsupervised anomaly detection}
To assess the quality of the flow representations in our model we devise an anomaly detection process that is based on flow similarity. Specifically, we compare all flows from the testing dataset $f_{\text{eval}}$ with the benign flows from the training dataset $f_{\text{train}}$ by calculating the cosine similarity for each possible pair, as illustrated in Equation \ref{similarity_score}, using the flow representation output of our model. From these we take the cosine similarity of the pair with the highest similarity as the similarity score between a flow and the benign traffic of the training dataset. Since the training data do not contain malicious flows,we expect that a malicious flow will have a low similarity score to flows of the training dataset, while a benign flow will most likely have high similarity to at least one flow from the training dataset. 

\begin{equation} \label{similarity_score}
\text{sim}(f_{\text{eval}},f_{\text{train}}) = \max \left( \frac{f_{\text{eval}}^T f_{train}}{\| f_{\text{eval}} \| \| f_{train} \|} \right)
\end{equation}

\begin{table}[t] 
    \centering
    \caption{Supervised evaluation for contrastive DNN on Netflows}
    \label{table:supervised_dnn}
    \resizebox{\columnwidth}{!}{ 
    \begin{tabular}{|c|c|c|c|c|} 
     \hline
     \textbf{\backslashbox{Train}{Test}} & \textbf{CICIDS} & \textbf{UNSW-NB}  & \textbf{CTU}  & \textbf{CICDDOS} \\ [0.3ex] 
     \hline
     \textbf{CICIDS} & \textbf{99\%} & 60\% & \textbf{67\%} & 52\%\\
     \hline
     \textbf{UNSW-NB} & 62\% & \textbf{98\%} & 54\% & 51\%\\ 
     \hline
     \textbf{CTU} & 77\% & 55\% & \textbf{98\%} & 48\%\\ 
     \hline
     \textbf{CICDDOS} & 85\% & 51\% & 50\% & \textbf{97\%} \\
     \hline
    \end{tabular}
    }
    \label{tab:bigtable}
\end{table}

In Tables \ref{table:unsupervised_transformer}, \ref{table:unsupervised_dnn}, \ref{table:supervised_dnn} we present the results of the evaluation within intra and inter-dataset settings for all baselines. The highest score for each case is denoted with bold numbers. The performance of self-supervised models to identify anomalies in an environment where information about ordinary benign traffic is available is evaluated in the intra-dataset evaluation, where benign flows are split between the testing and training datasets. In the intra-dataset evaluation we observe that supervised training with NetFlows outperforms both our self-supervised transformer with packet sequences and DNN baselines with Netflows, by up to 4\% in the case of the CICDDOS dataset when compared to the self-supervised transformer-based model. This is expected as this scenario is the least challenging case of intrusion detection where the training and evaluation dataset are fairly similar, thus supervised learning is sufficient to achieve good results. When comparing the self-supervised approaches on the intra-dataset evaluation, the transformer-based model comes on top with up to 3\% higher AUC score on the CTU dataset compared to the DNN model that employs Netflows. 

For the inter-dataset scenarios our model surpasses the self-supervised and supervised Netflow baselines in almost all cases, having up to 20\% higher AUC score compared to the self-supervised Netflow baseline when training on the UNSW-NB dataset and evaluating on the CTU dataset and vice versa. The only scenario where our approach does not have the lead in inter-dataset is on the scenarrio where we train on the CICIDS dataset and evaluate on the CTU dataset, where the supervised Netflow approach has 4\% higher AUC score than our approach. However, our model still shows a 10\% higher AUC score than the self-supervised model using Netflows in this case. In the inter-dataset evaluation, benign flows in the training and testing datasets are derived from distinct environments. Since benign flows can differ vastly from one domain to another, it becomes easier for self-supervised models to mistakenly identify benign flows in the testing dataset as malicious when comparing them to those of the training dataset. Consequently, the evaluation becomes harder. The challenging aspect of this task becomes apparent from the supervised baseline which has low AUC scores for all inter-dataset evaluations, while the self-supervised models show decently high AUC scores with our transformer-based model demonstrating the best performance once more. 

These results verify that the self-supervised learning process, in conjunction with our transformer-based model architecture, has indeed acquired the ability to detect similarities and differences between flows, enabling the model to provide rich feature representations for flow packet sequences that can be used without the need for supervised training.

\begin{table}[t]
\centering
\caption{AUC scores for few-shot supervised finetuning for contrastive transformers on packet sequences}
\label{table:auc_supervised_bert}
\resizebox{\columnwidth}{!}{ 
\begin{tabular}{|c|c|c|c|} 
 \hline
\textbf{\backslashbox{Pre-Train}{Fine-tune}} & \textbf{Random weights} & \textbf{Pretrained} \\ [0.3ex] 
 \hline
 \textbf{CICIDS} & 96.9\% & 99.4\% \\
 \hline
 \textbf{UNSW-NB} & 98.3\% & 99.2\% \\ 
 \hline
 \textbf{CTU} & 94.2\% & 96.3\% \\ 
 \hline
 \textbf{CICDDOS} & 91\% & 93.4\% \\ 
 \hline
\end{tabular}
}
\end{table}

\begin{table}[t]
\centering
\caption{AUC scores for few-shot supervised finetuning for contrastive DNN on Netflows}
\label{table:auc_supervised_dnn}
\resizebox{\columnwidth}{!}{ 
\begin{tabular}{|c|c|c|c|} 
 \hline
\textbf{\backslashbox{Pre-Train}{Fine-tune}} & \textbf{Random Weights} & \textbf{Pretrained} \\ [0.3ex] 
 \hline
 \textbf{CICIDS} & 97.2\% & 98.4\% \\
 \hline
 \textbf{UNSW-NB} & 98.6\% & 98.9\% \\ 
 \hline
 \textbf{CTU} & 93.8\% & 95.3\% \\ 
 \hline
 \textbf{CICDDOS} & 91.3\% & 91.9\% \\ 
 \hline
\end{tabular}
}
\end{table} 

\subsection{Supervised finetuning evaluation}
In this subsection we assess the benefits of pretraining in cases where some labeled data are available for supervised learning. In addition to finetuning we also assess the feasibility of transfer learning through our pretraining procedure in intra-dataset and inter-dataset settings. Specifically, in this evaluation our model along with the self-supervised DNN baseline were fine-tuned with a small amount of benign and malicious data and we compared their performances with and without employing pretraining. We fine-tuned each model for a maximum of 30 epochs with early stopping with patience 3 on the classification error of the validation set. For our few-shot learning evaluation we used 0.1\% of all labeled data in each dataset.

We can see in Tables \ref{table:auc_supervised_bert} and \ref{table:auc_supervised_dnn} that pretraining does increase the AUC score, regardless of the case increasing the AUC score up to 2.5\% using either Netflows or packet sequences. Although our model initially has a lower AUC score in certain instances, it is able to surpass the pretrained DNN baseline with Netflows with pretraining achieving up to 1.5\% when pre-training and fine-tuning on the CICDDOS dataset. The pretraining procedure can provide us with a superior model as a starting point when labeled data and, particularly, malicious flow samples are scarce.

Lastly, Table \ref{table:bert_transfer_learning} presents the results of our model's ability to facilitate transfer learning between datasets. In this scenario each model is pretrained on the benign flows of a single dataset and subsequently fine-tuned using a small amount of labeled data from a target dataset. From this process, it is evident that there is still an improvement in the AUC scores of the classifier in comparison to the scores obtained with randomly initialized weights. This demonstrates that the knowledge acquired through the pretraining process can be transferred from the domain of one dataset to another. Additionally, in comparison to the respective results of the self-supervised DNN model using Netflows in Table \ref{table:dnn_transfer_learning} our model also shows improved or equal performance for inter-dataset transfer learning with an improvement of up to 0.9\% in the case of pre-training on the UNSW-NB and fine-tuning on the CICIDS dataset.





\begin{table}[t]
\centering
\caption{AUC scores for transfer learning performance for contrastive transformers on packet sequences}
\label{table:bert_transfer_learning}
\resizebox{\columnwidth}{!}{ 
\begin{tabular}{|c|c|c|c|c|} 
\hline
\textbf{\backslashbox{Pre-Train}{Fine-tune}} & \textbf{CICIDS} & \textbf{UNSW-NB} & \textbf{CTU} & \textbf{CICDDOS}\\  
\hline
\textbf{CICIDS} & 99.4\% & 99\% & 94.9\% & 92.1\% \\ 
\hline
\textbf{UNSW-NB} & 99.1\% & 99.2\% & 94.9\% & 91.7\% \\ 
\hline
\textbf{CTU} & 97.8\% & 98.8\% & 96.3\% & 91.9\% \\ 
\hline
\textbf{CICDDOS} & 98.5\% & 98.7\% & 94.4\% & 93.4\% \\ 
\hline
\end{tabular}
}
\end{table}

\begin{table}[t]
\centering
\caption{AUC scores for transfer learning performance for contrastive DNN on Netflows}
\label{table:dnn_transfer_learning}
\resizebox{\columnwidth}{!}{ 
\begin{tabular}{|c|c|c|c|c|} 
 \hline
\textbf{\backslashbox{Pre-Train}{Fine-tune}} & \textbf{CICIDS} & \textbf{UNSW-NB} & \textbf{CTU} & \textbf{CICDDOS}\\ [0.3ex] 
 \hline
 \textbf{CICIDS} & 98.4\% & 98.8\% & 94.2\% & 91.8\% \\ 
 \hline
 \textbf{UNSW-NB} & 98.2\% & 98.9\% & 94.7\% & 91.3\% \\ 
  \hline
 \textbf{CTU} & 97.5\% & 98.8\% & 95.3\% & 91.6\% \\ 
  \hline
 \textbf{CICDDOS} & 98.1\% & 98.5\% & 94\% & 91.9\% \\ 

 \hline
\end{tabular}
}
\end{table}

\section{Conclusions}
In this paper we proposed a transformed-based model that utilizes self-supervised contrastive learning to enable generalizable intrusion detection. Our model is capable of directly processing sequences of packets to provide a flow representation that can be leveraged to identify anomalies in network traffic or to enhance supervised learning in scenarios where labeled traffic is either unavailable or available in limited quantities. The proposed contrastive learning approach employs packet replacement to create unique sequences for the pretraining task, which enables our model to learn and identify similarities or differences in the granularity of the packet between flows. Our approach demonstrates an improvement over similar self-supervised models on NetFlow datasets, both in supervised and unsupervised evaluation as a result of our transformer-based architecture, showing the effectiveness of our model to adapt to benign and malicious traffic from different domains.

\bibliographystyle{IEEEtran}
\bibliography{refs}

\end{document}